\documentclass{sn-jnl}
\usepackage{amsmath}
\usepackage{amssymb}
\usepackage{graphicx}
\usepackage{hyperref}
\usepackage{nicefrac}

\newcommand{\He}{$^3$He}
\newcommand{\Hea}{$^3$He-A}
\newcommand{\Heaero}{$^3$He-Aerogel}
\newcommand{\ute}{UTe$_{2}$}
\newcommand{\upt}{UPt$_{3}$}

\newcommand{\sro}{Sr$_{2}$RuO$_{4}$}
\newcommand{\kb}{k_{\text{B}}}
\newcommand{\grad}{{\vec\nabla}}
\def\dyadic#1{\overset{\mbox{$\leftrightarrow$}}{\large #1}}
\def\Tr#1{\mbox{Tr}\big\{#1\big\}}
\renewcommand{\vec}[1]{{\pmb #1}}
\newcommand{\whtaux}{{\widehat{\tau}_1}}
\newcommand{\whtauy}{{\widehat{\tau}_2}}
\newcommand{\whtauz}{{\widehat{\tau}_3}}
\def\ns{\negthickspace}

\begin{document}
\title{Anomalous Thermal Hall Effect in Chiral Phases of $^3$He-Aerogel}
\author*{Priya Sharma$^1$}
\affil{$^1$Department of Physics, Royal Holloway University of London \\ Egham, Surrey, TW20 0EX, UK}
\author{J. A. Sauls$^2$}
\affil{$^2$Department of Physics, Northwestern University \\ Evanston, 60208, Illinois, USA}
\abstract{
We report theoretical results for heat transport by quasiparticle excitations in superfluid $^3$He infused into silica aerogel engineered with uniaxial anisotropy. For this system two distinct equal spin pairing (ESP) superfluid phases have been reported based on NMR spectroscopy. Theoretical analysis predicts the first ESP state to be the chiral A phase with chiral axis aligned along the strain axis, and the lower temperature phase to be a polar-distorted chiral phase with random transverse fluctuations in the orientation of the chiral axis. We report calculations of heat transport for the high temperature chiral phase, including an anomalous (zero field) thermal Hall current originating from branch conversion scattering of Bogoliubov quasiparticles by the chiral order parameter induced by potential scattering by the silica aerogel. Observation of an anomalous thermal Hall current would provide a direct signature of the underlying chirality and topology of the superfluid phase of $^3$He in ``stretched'' silica aerogels.
}


\maketitle

\vspace*{-5mm}
\subsection*{Introduction}

There are two stable superfluid phases of liquid $^3$He that are realisations of nontrivial topological superfluid phases. The A-phase corresponds to the Anderson-Brinkman-Morel (ABM) state~\cite{and61}, which spontaneously breaks mirror and time-reversal symmetries, and is also a topologial superfluid~\cite{vol92,she16}. The B-phase corresponds to the Balian-Werthamer state~\cite{bal63}, and is an isotropic three-dimensional time-reversal-invariant topological phase~\cite{chu09,nag09,vol09,miz12,wu13}. While pure bulk liquid $^3$He is intrinsically pristine, $^3$He in a random potential is realised when $^3$He is infused into highly porous silica aerogel~\cite{por95,spr95}. The $^3$He-aerogel system has been investigated extensively as new superfluid phases are discovered in this system~\cite{hal18,hal19}, offering opportunities to study novel effects of disorder on the superfluids such as impurity-induced quantum phase transitions~\cite{mat97}, nematic superfluid phases~\cite{ask12} 
and half-quantum vortices~\cite{aut16,reg21}.

Aerogels are characterised by their porosity and degree of anisotropy. While silica aerogels can be synthesised with good control over their porosity, advances in aerogel synthesis techniques have led to the engineering of ``stretched'' and compressed silica aerogels with characterisable uniaxial anisotropies~\cite{pol08}. Nuclear magnetic resonance (NMR) experiments on $^3$He infused into uniformly anisotropic, axially stretched silica aerogel report the stabilisation of an ESP chiral state~\cite{pol12}. At lower temperatures, a sharp transition to a second ESP phase is observed. Theoretical analysis based on Ginzburg-Landau theory predicts a transition from the normal state to a chiral ABM state with the chirality axis aligned along the strain axis of aerogel anisotropy~\cite{sau13}. A second transition is predicted at a lower temperature to a \emph{biaxial state}, which is also chiral, with an order parameter that also breaks axial rotation symmetry. Random anisotropy of the aerogel is then predicted to generate an inhomogeneous Larkin-Imry-Ma (LIM) state with reduced chiral order along the anisotropy axis of the aerogel~\cite{sau13}.
In another class of highly anisotropic, high-porosity \emph{Nafen} aerogels, a third superfluid phase - the ``polar phase'' - has been observed with a single p-wave orbital aligned along the Nafen anisotropy axis~\cite{dmi15}. This phase is favored in globally uniaxial aerogels~\cite{aoy06,reg21}, and is also theoretically predicted to be stabilised when $^3$He is confined in cylindrical pores of diameter $\sim 100\,\mbox{nm}$~\cite{wim13}. The discovery of the polar phase of $^3$He infused into Nafen aerogel was the catalyst for the discovery of half-quantum vortices~\cite{aut16}.

Andreev scattering at the confining surfaces in a superfluid with broken mirror and time-reversal symmetries generates chiral fermions confined on edge boundaries. These states carry uni-directional edge currents which are the source of the ground-state angular momentum~\cite{sau11}. Chiral edge states reflect the broken mirror and time-reversal symmetry and non-trivial topology of the bulk ground state. In the presence of a thermal bias between counter-propagating currents on opposing boundaries, the edge states give rise to an anomalous (zero external field) thermal Hall conductance in $d=2$ which is quantized for edges with local translational invariance, $K_{xy}^{\mbox{\tiny edge}}/T=\pi\kb^2/6\hbar$~\cite{rea00,gos15}, and for which the chiral axis - perpendicular to the plane containing the edge currents - plays the role of the magnetic field axis.
Impurities embedded into a chiral superconductor also support chiral currents bound to each impurity. This impurity spectrum, combined with potential scattering of unbound quasiparticles, generates resonant skew scattering of a nonequilibrium population of quasiparticles. This mechanism is responsible for \emph{bulk anomalous Hall effects}~\cite{she16,nga20}. In contrast to the quantized edge current for idealized boundaries, the impurity-induced anomalous Hall effect depends on the impurity cross-section, Fermi surface topology and Chern number of the chiral superconductor. Thus, the observation of an anomalous thermal Hall current can be used as direct detection of broken mirror and time-reversal symmetries, and can also reveal the Chern number in putative topological superconductors such as $UPt_3$, $UTe_2$, $Sr_2RuO_4$~\cite{nga20}.
In this paper, we report theoretical calculations of anomalous thermal Hall transport for the ABM phase of superfluid $^3$He confined in anisotropic silica aerogels. Observation of an anomalous thermal Hall effect in superfluid $^3$He infused into anisotropic aerogels would provide direct evidence of the theoretical prediction~\cite{sau13} that the ESP state identified by NMR~\cite{pol12} in stretched silia aerogel is the chiral ABM state.

The linear response of a system to a weak perturbation, in this case a thermal bias or local temperature gradient, depends on the equilibrium state of the system, particularly the spectral function for quasiparticles and Cooper pairs, as well as the distribution of excitations. For systems with quenched disorder, i.e. \He\ infused into silica aerogel, additional statistical averaging is required.     

\vspace*{-3mm}
\subsection*{Impurity Scattering Model for $^3$He in Aerogel}

High-porosity (e.g. 98\%) silica aerogels are random solids with fractal correlations composed of silica clusters and strands of typical diameter $d\simeq 3\,\mbox{nm}$, with most probable strand separation $\xi_a\sim 40\,\mbox{nm}$, also referred to as the aerogel correlation length. The size of a typical cluster or strand is large compared to the Fermi wavelength, $k_f^{-1}\sim 0.1\,\mbox{nm}$, but is small compared to the coherence length, $\xi_0\simeq 20-80\,\mbox{nm}$ over the full pressure range, which is the size of Cooper pairs. Thus, the simplest model for $^3$He infused into aerogel is to treat the aerogel as a coarse-grained distribution of impurities that scatter otherwise ballistic $^3$He quasiparticles. In the limit $\xi_0 \gg \xi_a$ then \He\ in aerogel is well described by a \emph{homogeneous}, isotropic scattering medium  (IHSM) with a mean-free path determined by the structure of aerogel~\cite{thu98}. This limit is achieved at low pressures near the quantum critical pressure, but at high pressures the structure correlations of the aerogel become relevant~\cite{sau05}.   
Here, we consider the simplest IHSM which is parametrized by a mean density of scattering impurities, $\bar{n}_s$, and scattering cross-section, $\sigma$, for point-like impurities, and we neglect the effects of aerogel correlations. The latter can be included if refined predictions are needed for comparison with future experiments.

The equilibrium retarted (R) and advanced (A) propagators for quasiparticles and Cooper pairs in \Heaero\ obey Eilenberger's quasiclassical transport equations~\cite{eil68},
\begin{equation}\label{QCeqn}
\left[\varepsilon\whtauz - \widehat{\Sigma}^{R,A} - \widehat{\Delta}\,,\,\widehat{g}^{R,A}\right] 
+
i\vec{v}_\vec{p}\cdot\grad\widehat{g}^{R,A} = 0 
\end{equation}
and the normalization conditions,
\vspace*{-3mm}
\begin{equation}\label{normalization}
\left[\widehat{g}^{R,A}\right]^2 = -\pi^2\widehat{1}
\,,
\vspace*{-3mm}
\end{equation}
where $\widehat{\tau}_{j}$ $(j = 1,2,3)$ are the Pauli matrices and $\widehat{1}$ is the corresponding unit matrix in Nambu space, $\widehat{\Delta}$ is the mean-field pairing self energy (order parameter) and $\widehat{\Sigma}^{R,A}$ is the impurity self-energy, which in general renormalizes both the quasiparticle energies and pairing self energy. For homogeneous equilibrium, coarse-grain averaged over the impurity distribution, the equilibrium propagators reduce to
\vspace*{-5mm}
\begin{equation}\label{eq-gRA_equilibrium}
\widehat{g}^{R,A}(\hat{\vec{p}},\varepsilon) 
=
-\pi\frac{\tilde{\varepsilon}^{R,A}(\hat{\vec{p}},\varepsilon)\whtauz-
                 \widehat{\tilde\Delta}(\hat{\vec{p}},\varepsilon)}
         {\sqrt{\vert\tilde\Delta(\hat{\vec{p}},\varepsilon)\vert^2-
            (\tilde\varepsilon^{R,A}(\hat{\vec{p}},\varepsilon))^2}}
\,,
\vspace*{-3mm}
\end{equation}
\begin{equation}
\hspace*{-10mm}\mbox{where} 
\quad
\tilde\varepsilon^{R,A}=\varepsilon\pm i0^{+}-\Sigma_{\mbox{\tiny imp}}^{R,A}(\hat{\vec{p}},\varepsilon)
\,,
\mbox{and}\quad
\tilde\Delta^{R,A} = \Delta(\hat{\vec{p}}) + \Delta_{\mbox{\tiny imp}}^{R,A}(\hat{\vec{p}},\varepsilon)
\,,
\end{equation}

\noindent are the impurity renormalized excitation spectrum and order parameter. 
The Nambu representation of the mean field order parameter for the chiral p-wave ESP state with $\vec\ell=\hat{\vec{z}}$ and $\hat{\vec{d}}=\hat{\vec{z}}$ is 
\vspace*{-3mm}
\begin{equation}\label{eq-ABM_OP}
\widehat\Delta(\hat{\vec{p}})
=
\frac{\Delta}{\sqrt{2}}\,
\begin{pmatrix}
0	&	\sigma_x\,\sin\theta_{\hat{\vec{p}}}\,e^{+i\phi_{\hat{\vec{p}}}}
\cr
\sigma_x\,\sin\theta_{\hat{\vec{p}}}\,e^{-i\phi_{\hat{\vec{p}}}}	&	0
\end{pmatrix}
=
\frac{\Delta}{\sqrt{2}}\,\left(\sigma_x\hat{\vec{p}}_x\whtaux - \sigma_x\hat{\vec{p}}_y\whtauy\right)
\,.
\vspace*{-2mm}
\end{equation}

\noindent For the IHSM of \He\ infused into silica aerogel the off-diagonal impurity self energy vanishes, $\widehat{\Delta}_{\mbox{\tiny imp}} \equiv 0$; this result follows for any unconventional order parameter for which $\langle\widehat\Delta(\hat{\vec{p}})\rangle_{\hat{\vec{p}}} =0$. The pairing self energy is then determined by the BCS gap equation,
\begin{equation}
\Delta(\hat{\vec{p}})=
\int^{+\varepsilon_c}_{-\varepsilon_c}\ns\ns
\frac{d\varepsilon}{4\pi i}\,
\int d^2\vec{p}'
\lambda(\hat{\vec{p}}\cdot\hat{\vec{p}}')
\mbox{\sf Im}
\frac{\Delta(\hat{\vec{p}}')}
{\sqrt{\mid{\Delta(\hat{\vec{p}}')}\mid^2-\tilde{\varepsilon}^{R}(\hat{\vec{p}}',\varepsilon)^2}}
\times\tanh\left(\frac{\varepsilon}{2T}\right)
\,,
\end{equation} 
where $\lambda(\hat{\vec{p}},\hat{\vec{p}}')=3\lambda_1(\hat{\vec{p}}\cdot\hat{\vec{p}}')$ for p-wave pairing and $\int d^2\vec{p}'(\ldots)\equiv\langle(\ldots)\rangle_{\vec{p}'}$ is the average over the Fermi surface. The cutoff $\varepsilon_c$ and pairing interaction $\lambda_1$ are eliminated in favour of $T_c$ for the transition in pure bulk \He\ using the linearized gap equation. Scattering by the aerogel suppresses the transition temperature depending on the scattering cross section and the magnitidue of the impurity density~\cite{thu98}.
The cross section for quasiparticle-aerogel scattering is obtained from T-matrix (retarded 'R' and advanced 'A'),
%
\begin{eqnarray}\label{t-matrix-eqn}
\widehat{T}^{R,A}(\vec{p},\vec{p}';\ns\varepsilon) 
\ns=\ns 
\widehat{U}(\vec{p},\vec{p}')
\ns+\ns 
N_f\int\ns d\vec{p}''\,\widehat{U}(\vec{p},\vec{p}'')
\,\widehat{g}^{R,A}(\vec{p}'',\varepsilon)\,\widehat{T}^{R,A}(\vec{p}'',\vec{p}';\varepsilon)
\,.
\end{eqnarray}
The T-matrix describes multiple scattering of quasiparticles by the impurity potential. The main subtlety is that the Bethe-Salpeter equation depends on the exact propagator, $\widehat{g}^{R,A}$, including the self-energy corrections from the scattering medium, which must be calculated self-consistently with the solution for the T-matrix. 

For \emph{normal} \He\ in aerogel described by the IHSM, $\widehat{U} = U_0\widehat{1}$ is the strength of the s-wave scattering potential. Thus, the normal-state propagator in equilibrium simplifies to $\widehat{g}^{R,A}_{\mbox{\tiny N}}=\mp i\pi\whtauz$, with the result 
\begin{equation}
\widehat{T}_{\mbox{\tiny N}}^{R/A}=\frac{1}{\pi N_f}\,\sin\delta_0\,e^{\mp i\delta_0\whtauz}
\,,
\end{equation}
where $\delta_0$ is the s-wave scattering phase shift defined by $\delta_0=\tan^{-1}(\pi N_f U_0)$. In this IHSM, the mean density of impurities, $\bar{n}_s$, and scattering rate for normal-state quasiparticles, $\hbar/\tau_{\mbox{\tiny N}}$, are related to the mean free path $l=v_f\tau_{\mbox{\tiny N}}$ and scattering cross section 
\begin{equation}
\bar{n}_s = \frac{1}{\sigma\;l}
\quad\mbox{with}\quad
\sigma=\frac{4\pi\hbar^2}{p_f^2}\,\bar{\sigma}
\,,
\end{equation}
where the normalized cross section, $\bar\sigma$ is related to the scattering potential by
\begin{equation}
\bar\sigma = \frac{(\pi N_f U_0)^2}{1+(\pi N_f U_0)^2} = \sin^2\delta_0
\,,
\end{equation}
and ranges from $\bar\sigma\approx 0$ (weak scattering/Born limit) to $\bar\sigma\rightarrow 1$ (strong scattering/Unitarity limit). In the IHSM the corresponding impurity self-energy is given by the forward scattering limit of the T-matrix, which for s-wave scattering is simply,  
\begin{equation}\label{eq-SigmaRA_equilibrium}
\widehat{\Sigma}^{R,A,K} = \bar{n}_{s}\,\widehat{T}^{R,A,K} 
\,. 
\end{equation}
Thus, for normal-state quasiparticles in equilibrium we obtain,
\begin{equation}
\widehat{\Sigma}^{R,A}_{\mbox{\tiny N}}
=
\frac{\bar{n}_{s}}{\pi N_f}\,\sin\delta_0\cos\delta_0\,\widehat{1}
\mp i
\frac{\bar{n}_{s}}{\pi N_f}\,\sin^2\delta_0\,\whtauz
\,.
\end{equation}
The unit matrix term drops out of the transport equation and does not contribute to static physical properties. The $\whtauz$ terms is purely imaginary and represents the scattering rate of normal-state quasiparticles by the medium of impurites,
\begin{equation}
\frac{\hbar}{\tau_{\mbox{\tiny N}}} = \frac{\bar{n}_s}{\pi N_f}\sin^2\delta_0
\,,
\end{equation}
and corresponding mean-free path, $l=v_f\tau_{\mbox{\tiny N}}$. In what follows we use the parameters, $l$ and $\bar\sigma$ to characterize aerogel scattering.

\begin{figure}[]
\centering
\includegraphics[height=2in,width=2.3in,angle=0]{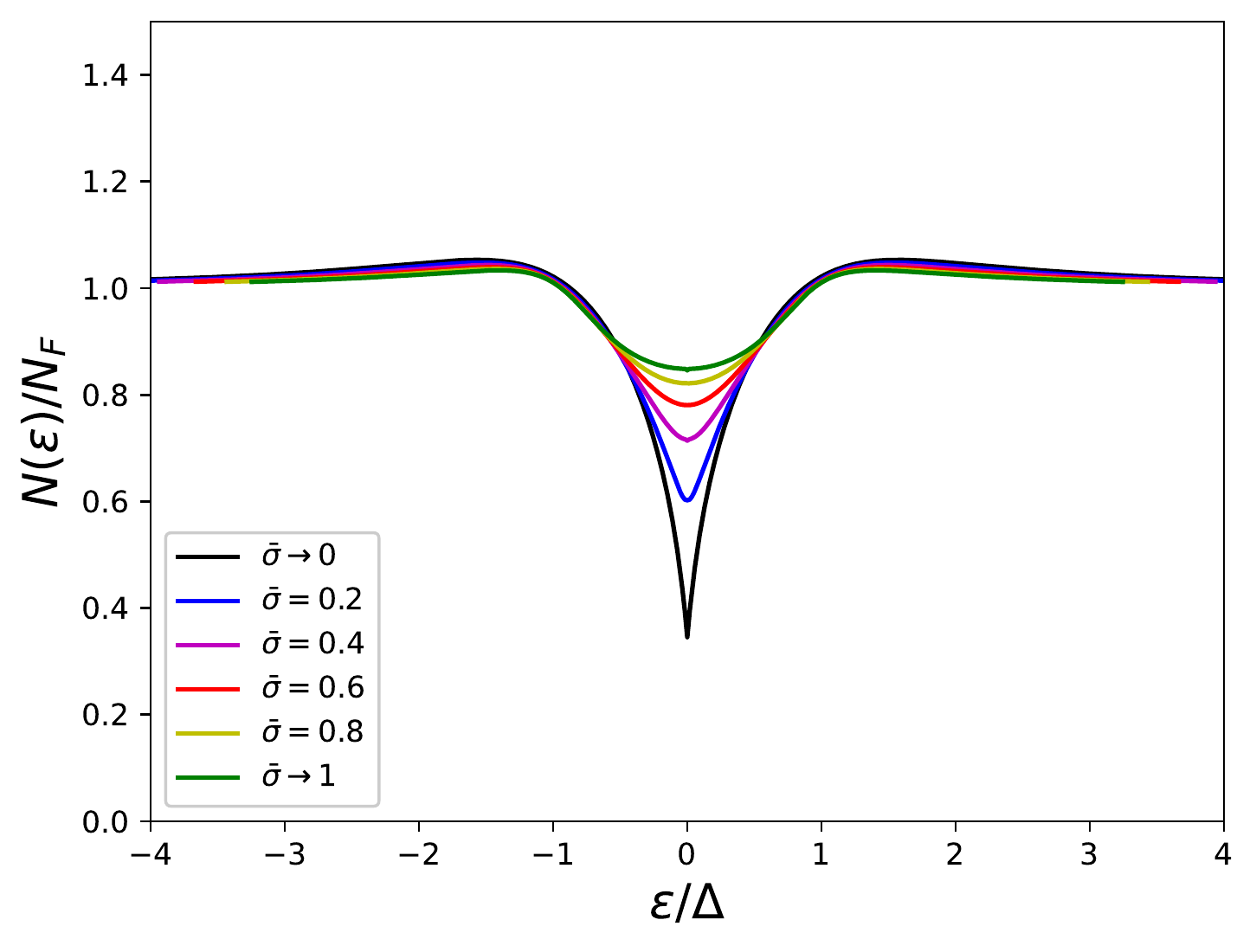}
\includegraphics[height=2in,width=2.3in,angle=0]{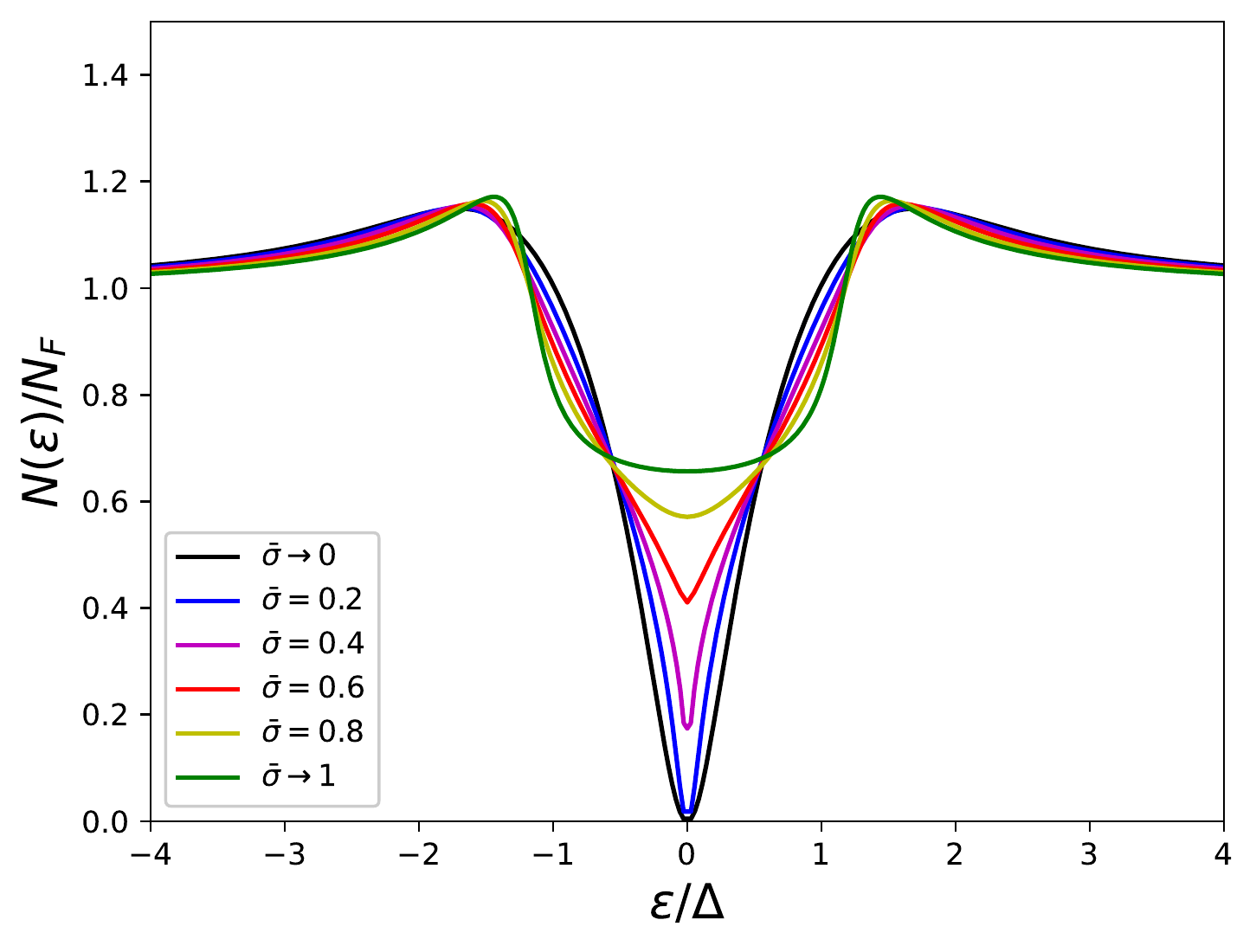}
\caption{Density of states for the chiral ABM state of $^3$He in aerogel with $\ell=113\,\mbox{nm}$ at pressures $p=14.1\,\mbox{bar}$ (left panel) and $p=26\,\mbox{bar}$ (right panel) for reduced temperature $T=0.2 T_c$ over a range of impurity cross sections.}
\label{figAphaseDos}
\end{figure}

Below the superfluid transition impurity scattering leads to pair breaking, suppression of the $T_c$ and the order parameter, and to the formation of a sub-gap quasiparticle spectrum. The latter plays a key role in the low temperature transport properties, and is visible in the quasiparticle density of states (DOS),
$N(\varepsilon)=N_f\int d^2\vec{p}\,\left[-\frac{1}{\pi}\mbox{\sf Im}\,g^R(\vec{p},\varepsilon)\right]$,
in Fig.~\ref{figAphaseDos} for a mean free path of $l=113\,\mbox{nm}$, pressures of $p=14.1\,\mbox{bar}$ and $p=26.0\,\mbox{bar}$ and as a function of the dimensionless cross section. Note that the superfluid becomes gapless away from the Born limit, with a large DOS at the Fermi level in the unitarity limit.

The equilibrium solutions for $\widehat{g}^{R,A}$, $\widehat\Sigma^{R,A}$ and $\widehat\Delta$ are essential for calculating the linear response functions. The resulting solutions for the nonequilibrium propagator, $\delta\widehat{g}^{R,A,K}$, the corresponding T-matrix, $\widehat{T}^{R,A,K}$ and self-energy, $\delta\widehat\Sigma^{R,A,K}$, are calculated and reported in the Appendix. These functions are the key components for calcuating the nonequilibrium heat current. In what follows we summarize our results for the longitudinal thermal conductivity and the anomalous Hall current.  

\subsection*{Heat Transport in the Chiral A phase in Aerogel}

For a thermal gradient, $\grad T$, the linear response relation for the heat current density in the absence of a mass transport current is
\begin{equation}
\vec{j}_{\varepsilon} = - \dyadic{K}\cdot\grad T
\,,\quad \mbox{where}\quad\,
\dyadic{K} \begin{pmatrix} 
		K_{xx} & K_{xy} & 0 \cr -K_{xy} & K_{xx} & 0 \cr 0 & 0 & K_{zz}
           \end{pmatrix} 
\,,
\end{equation}
is the thermal conductivity tensor allowed by chiral symmetry. Here we consider an in-plane thermal gradient, e.g. $\grad T = (\partial T/\partial x)\hat{\vec{x}}$, in which case the longitudinal heat current is 
$j_x = -K_{xx}\,(\partial T/\partial x)$ and the transverse ``Hall'' current is $j_y =-K_{xy}(\partial T/\partial x)$.   

In order to calculate the conductivity tensor we use Keldysh's formulation~\cite{kel65} of nonequilibrium quasiclassical theory~\cite{eil68,lar69,ser83,rai94b}. 
This theory extends the microscopic formulation of Fermi liquid theory to include Cooper pair formation and condensation. The theory describes equilibrium and nonequilibrium phenomena over length scales larger than the Fermi wavelength, $k_f^{-1}$, and time scales longer than the inverse Fermi energy $\hbar/E_f$. We follow the formalism of Ref.~\cite{gra96a}, and include additional technical steps related to this analysis in the Appendix. In particular, the heat current density can be calculated from the Keldysh propagator, $\widehat{g}^K$,

\begin{equation}\label{je}
\vec{j}_{\varepsilon}= 
N_f
\int d^2\vec{p}
\int\frac{d\varepsilon}{4\pi i}
\left[\varepsilon\,\vec{v}_{\hat{\vec{p}}}\right]\,\Tr{\widehat{g}^K(\vec{p},\varepsilon)}
\,,
\end{equation}
where $\varepsilon\,\vec{v}_{\hat{\vec{p}}}$ is the heat current carried by a normal-state quasiparticle, $N_f$ is the normal-state density of states at the Fermi level. The integration is an average over all trajectories defined by momenta on the Fermi surface, $\vec{p}=p_f\hat{\vec{p}}$, which for $^3$He are the unit vectors normal to the Fermi sphere, $\hat{\vec{p}}$, i.e. an angular integral $\int\,d^2\vec{p}(\ldots)$, combined with integration over the spectrum and trace over the distribution functions for the branches of nonequilibrium Bogoliubov quasiparticles. The spectral function, nonequilibrium distribution function and coherence factor for heat transport are encoded in the nonequilibrium Keldysh propagator, $\widehat{g}^K(\hat{\vec{p}},\varepsilon)$; the calculation of which is outlined in the Appendix.

Theoretical results for the longitudinal heat current of \He\ based on the IHSM were reported by the authors in Ref.~\cite{sha03}. However, we neglected the nonequilibrium correction to the off-diagonal impurity self energy, i.e. ``vertex corrections'' in the diagrammatic formulation of the Kubo response function.  
Since the perturbation, $\propto\vec{v}_{\hat{\vec{p}}}\cdot\nabla T$, has $p$-wave ($l=1$) symmetry; and the off-diagonal propagator and self-energies belong to the same ($l=1$) channel, the vertex correction is non-zero even within the IHSM, where impurity scattering is isotropic (i.e. $l=0$ channel only)~\cite{nga20}. 
In particular, 
\begin{eqnarray}
K_{ij}
=
(K^d_{xx} + K^a_{xx})
(\delta_{ij}-\hat{\vec{z}}_i\hat{\vec{z}}_j)
+
(K^d_{zz} + K^a_{zz})
\hat{\vec{z}}_i\hat{\vec{z}}_j
+ 
K^a_{xy}
(
\hat{\vec{x}}_i\hat{\vec{y}}_j
-
\hat{\vec{y}}_i\hat{\vec{x}}_j
)
\,, 
\end{eqnarray}
where $K_{ij}^d$ are the diagonal contributions to the thermal conductivity omitting the vertex corrections. These terms and the vertex corrections, $K^a_{ij}$, 
\begin{eqnarray}\label{Kaij}
K_{\{{{xx}\atop{xy}}\}}^a
&=&
\left\{\mp\right\}
\Gamma_N\frac{N_f\,v_f^2\,\Delta^2}{4\pi\,T^2}
\int\ns d\varepsilon\,\varepsilon^2\,\mbox{\sf sech}^2\left(\frac{\varepsilon}{2T}\right)
\nonumber\\
&&
\times\,{\mathsf Im}[\tilde{\varepsilon}^R]^2
\mathcal{B}^2(\varepsilon)\,
{\left\{
{\mbox{\sf Re}}\atop{\mbox{\sf Im}}
\right\}}
\left[\frac{\bar{T}_+^R[\bar{T}_-^R]^{\star}}
                   {1+\Gamma_N\,\bar{T}_+^R[\bar{T}_-^R]^{\star}\mathcal{A}(\varepsilon)}\right]
\,,
\end{eqnarray}
are discussed in detail in the Appendix. In Eq.~\eqref{Kaij} $\Delta$ is the amplude of the A-phase order parameter, and $\tilde{\varepsilon}^R$, the Fermi surface averages for $\mathcal{A}(\varepsilon)$ and $\mathcal{B}(\varepsilon)$, and the normalized T-matrix amplitudes, $\bar{T}^{R}_{\pm}$ are defined in the Appendix. Note also that $\Gamma_{\mbox{\tiny N}}=\frac{\bar{n}_{s}\bar\sigma}{\pi N_f}$ is the normal-state scattering rate in the aerogel. 

\begin{figure}[]
\centering
\includegraphics[width=0.5\columnwidth]{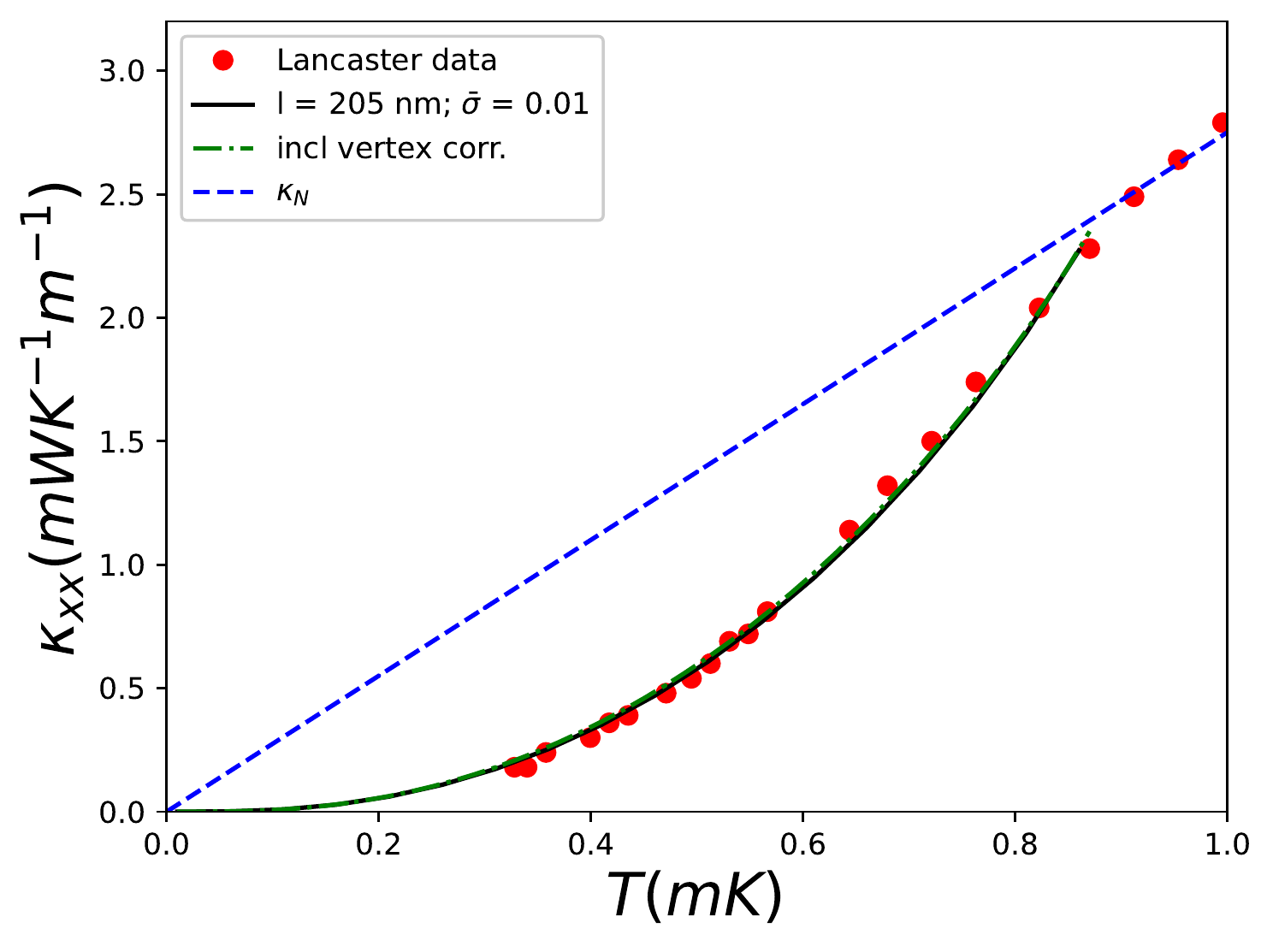}\includegraphics[width=0.5\columnwidth]{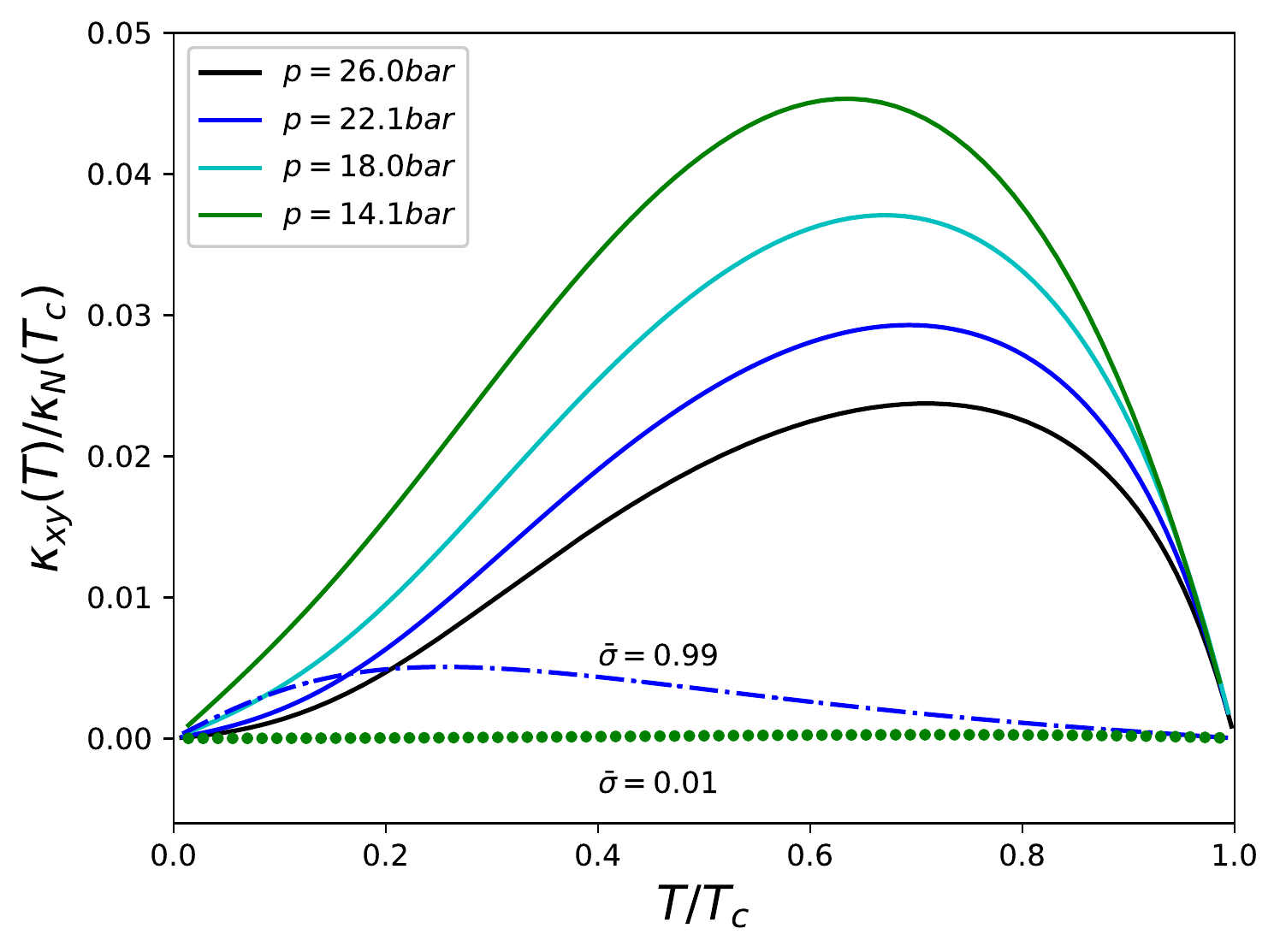}
\caption{
(a) Longitudinal conductivity $K_{xx}$ for \Hea\ in 98\% porous aerogel with $\vec{\ell}\parallel\hat{\vec{z}}$. The solid black line is the theoretical result for $l=205\,\mbox{nm}$ in the Born limit. The green dashed curve includes the vertex correction $K_{xx}^a$. Red circles are data from Ref.~\cite{fis02}.
(b) Hall conductivity $K_{xy}$ 
normalized to the normal-state conductivity at $T_c$. 
The linear onset of $K_{xy}$, maximum at intermediate temperature and suppression at low temperature are characteristic signatures of branch conversion scattering. 
The pressures and mean free path of $l=113\,\mbox{nm}$ correspond to the experiments reported in Ref.\cite{pol12} for the ESP state identified by NMR, and predicted in Ref.~\cite{sau13} to be the chiral ABM state with $\vec{\ell}\parallel\hat{\vec{z}}$. Solid lines correspond to $\bar\sigma=0.6$. Note the small magnitude of $K_{xy}$ compared to $K_{\mbox{\tiny N}}(T_c)$.
}
\label{fig-Kij}
\end{figure}

In the absence of vertex corrections, our result 
obtained in Appendix Eq.~\eqref{k-expression} 
reduces to the that obtained in Ref.~\cite{gra96a}. The vertex correction obtained here for the longitudinal conductivty, $K_{xx}^a$, is a small correction of order a few percent to the result for $K_{xx}^d$ of $^3$He-A reported in Ref.~\cite{sha03}, and barely resolvable as shown in Fig.~\ref{fig-Kij}(a).
However, the Hall conductivity is due entirely to the vertex correction, $K^a_{xy}$. It has a small prefactor when compared to the longitudinal conductivity at $T_c$ as shown in Fig.~\ref{fig-Kij}(b).

The physical mechanism responsible for the Hall conductivity in Eq.~\eqref{Kaij} is branch conversion scattering by the chiral order parameter induced by potential scattering in the presence of the thermal gradient. In branch conversion scattering a quasiparticle converts to a quasihole with the addition of a Cooper pair to the condensate, and thus acquires angular momentum of $-\hbar$, and similarly a quasihole can scatter into a quasiparticle, removing a Cooper pair and acquiring angular momentum $+\hbar$. The impurity potential and bound state spectrum also break particle-hole symmetry, as reflected by the term $\mbox{\sf Im}\tilde{\varepsilon}^R(\varepsilon)$, for any $0<\bar\sigma<1$, resulting in a net transfer of angular momentum between the condensate and the scattered quasiparticles, and as a result a transverse Hall current~\cite{nga20,she16}.
The conductivity is sensitive to the quasiparticle mean free path, $l$, and the pressure via the pair-breaking ratio, $\xi_0/l=\hbar v_f/2\pi \tau_{\mbox{\tiny N}} T_c$, and the dimensionless cross section, $\bar\sigma$, which reflects the degree of particle-hole asymmetry. In particular, impurity-induced particle-hole asymmetry vanishes in the Born limit, $\bar\sigma\rightarrow 0^+$, and also in the unitarity limit, $\bar\sigma=1$. This is evident in the vanishing of $K_{xy}$ in these two limits. The Hall conductivity $K_{xy}$ is maximum for $\bar\sigma\approx 0.6$, and peaks at $T/T_c\approx 0.6-0.7$. The maximum is suppressed and shifts to lower temperature as $\bar\sigma\rightarrow 1$
as shown in Fig.~\ref{fig-Kij}(b).

\subsection*{Conclusion and Outlook}

Anomalous (zero field) Hall transport is a signature of broken mirror and time-reversal symmetry in chiral condensates, and can be used as an experimental test for chirality of superfluid phases stabilised in anisotropic aerogels. The temperature dependence of the anomalous thermal Hall conductivity is a signature of branch conversion scattering by the chiral order parameter. The magnitude of the Hall conductivity is typically a few percent of the normal-state conductivity at $T_c$ for typical high porosity aerogels. The anomalous Hall conductivity is sensitive to the mean free path for quasiparticles set by the aerogel and to the particle-hole asymmetry induced by impurity scattering, the latter parametrised by the dimensionless scattering cross-section $0 \le \bar\sigma \le1$ in the IHSM for point impurities. Thus, the Hall conductivity vanishes in the limits where the impurity-induced particle-hole asymmetry vanishes viz, both Born ($\bar\sigma\rightarrow 0$) and unitarity ($\bar\sigma\rightarrow 1$) scattering limits. Finite size impurities ensure that particle-hole symmetry is broken by scattering in one of more channels~\cite{nga20}. We hope these quantitative predictions for Hall conductivity will motivate new experiments to test theoretical  predictions for chiral phases of \He\ in anisotropic aerogels. 

We expect anomalous thermal Hall transport to occur in superfluid \Hea\ confined in thin cavities for which scattering by an atomically rough surface plays the role of impurity scattering. In this case skew scattering by the top and bottom surfaces of the cavity will generate a component of the heat current in the plane parallel to the surface and transverse to the thermal gradient. 
In addition, the spectrum of gapless Weyl fermions confined along the edges of the cavity are predicted to generate a quantised anomalous thermal Hall conductivity~\cite{rea00,gos15}. An analysis of the relative importance of the surface and edge contributions to $K_{xy}$ for \Hea\ confined in thin cavities will be discussed in a separate report.
More broadly the impurity-induced anomalous thermal Hall effect can provide a powerful bulk transport probe to test for broken time-reversal and mirror symmetry in putative topological chiral superconductors such as \sro, \upt, and \ute.

\bmhead{Acknowledgments} 
The research of JAS was supported by US NSF Grant DMR-1508730. 
PS thanks Royal Holloway University of London for support. 
We thank Wave Ngampruetikorn for important discussions.

\begin{appendices}
\renewcommand\theequation{A.\arabic{equation}}
\subsection*{Appendix}

The nonequilibrium quasiclassical equation for the Keldysh propagator, $\widehat{g}^K$, is given by~\cite{gra96a}
\begin{equation}
\left[\varepsilon\widehat{\tau}_3-\widehat\Delta\,,\,\widehat{g}^K\right]_{\circ} 
+ 
i\vec{v}_{\hat{\vec{p}}}\cdot\vec{\nabla}\widehat{g}^K
-
\widehat{\Sigma}^R\circ\widehat{g}^K
+
\widehat{g}^K\circ\widehat{\Sigma}^A
-
\widehat{\Sigma}^K\circ\widehat{g}^A
+
\widehat{g}^R\circ\widehat{\Sigma}^K = 0
\,,
\end{equation}
where the convolution product is defined as,
\begin{equation}
\widehat{a}\circ\widehat{b}(\hat{\vec{p}},\vec{R},\varepsilon,\varepsilon')=
\int\frac{d\varepsilon''}{2\pi}
\widehat{a}(\hat{\vec{p}},\vec{R},\varepsilon,\varepsilon'')
\widehat{b}(\hat{\vec{p}},\vec{R},\varepsilon'',\varepsilon')
\end{equation}
The Keldysh propagator also satisfies the Keldysh extension of the normalization condition
\begin{equation}\label{gKcondition}
\widehat{g}^R\circ\widehat{g}^K + \widehat{g}^K\circ\widehat{g}^A = 0
\,.
\end{equation}
In the linear response regime the propagators $\widehat{g}^{R,A,K}$ can be expressed as perturbative corrections $\delta\widehat{g}^{R,A,K}$ to the equilibrium propagators external perturbations(thermal gradient in this case),
\begin{equation}\label{linearg}
\widehat{g}^{R,A,K} = \widehat{g}_0^{R,A,K} + \delta\widehat{g}^{R,A,K}
\,.
\end{equation}
The equilibrium quasiclassical propagators are given by Eq.~\eqref{eq-gRA_equilibrium} and self energies by 
Eq.~\eqref{eq-SigmaRA_equilibrium}; they also determine the equilibrium Keldysh propagator and self energy,
\begin{eqnarray}\label{phidefn}
\widehat{g}_0^K &=& \widehat{g}_0^R\,\phi_0 - \phi_0\,\widehat{g}_0^A
\\
\nonumber
\widehat{\Sigma}_0^K &=& \widehat{\Sigma}_0^R\,\phi_0 - \phi_0\,\widehat{\Sigma}_0^A
\,,
\end{eqnarray}
where $\phi_0(\varepsilon)=\tanh(\varepsilon/2T)$ is the equilibrium distribution function. For steady state nonequilibrium states considered here the convolution product simplifies to matrix multiplication in Nambu space.

The equations for the linear response functions simplify if we introduce the \emph{anomalous} propagator and self-energy~\cite{rai94b},
\begin{eqnarray}\label{anomalousdefn}
\delta\widehat{g}^K 
&=& 
\delta\widehat{g}^R\,\phi_0 - \phi_0\,\delta\widehat{g}^A + \delta\widehat{g}^a
\,,
\\
\delta\widehat{\Sigma}^K 
&=& 
\delta\widehat{\Sigma}^R\,\phi_0-\phi_0\,\delta\widehat{\Sigma}^A+\delta\widehat{\Sigma}^a
\,.
\end{eqnarray}
where $\phi_0$ is now the \emph{local} equilibrium distribution function parametrized by the local temperature, $T(\vec{R})$.
We assume that the local temperature is sufficiently slowly varying that we need retain only the gradient $\vec{\nabla}T$.
The resulting linear equations for $\delta\widehat{g}^{R,A,a}$ reduce to,
\begin{equation}\label{lineareqns}
\left[
\widehat{H}^{R,A}\,,\,\delta\widehat{g}^{R,A}
\right]
+i\vec{v}_{\hat{\vec{p}}}\cdot\vec{\nabla}\widehat{g}_0^{R,A}
= 
+ 
\left[
\widehat{g}_0^{R,A}\,,\,\delta\widehat{\Sigma}^{R,A}
\right]
\,,
\end{equation}
where 
\begin{equation}
\widehat{H}^{R,A} = \tilde{\varepsilon}^{R,A}\whtauz - \widehat{\Delta}(\hat{\vec{p}})
\,.
\end{equation}
For the anomalous propagator using Eqs.~\eqref{phidefn}, \eqref{anomalousdefn} and \eqref{lineareqns} we obtain,
\begin{eqnarray}\label{gaeqn}
\widehat{H}^R\,\delta\widehat{g}^a
- 
\delta\widehat{g}^a\,\widehat{H}^A  
+ 
i\vec{v}_{\hat{\vec{p}}}\cdot\vec{\nabla}\phi_0\,\left[\widehat{g}_0^R-\widehat{g}_0^A\right]
=
\delta\widehat{\Sigma}^a\,\widehat{g}_0^A
-
\widehat{g}_0^R\,\delta\widehat{\Sigma}^a
\,.
\end{eqnarray}
Thus, the driving perturbation is isolated on the left side of Eq.~\eqref{gaeqn} and the vertex corrections are
on the right side.
The linearized transport equations for $\delta\widehat{g}^{R,A,a}$ can be ``inverted'' efficiently using the normalization conditions expanded through linear order, 
\begin{eqnarray}\label{linearnormalization}
&\widehat{g}_0^{R,A}\,\delta\widehat{g}^{R,A} + \delta\widehat{g}^{R,A}\,\widehat{g}_0^{R,A} = 0&
\,,
\\
&\widehat{g}_0^R\,\delta\widehat{g}^a + \delta\widehat{g}^a\,\widehat{g}_0^A = 0&
\,,
\end{eqnarray}
and the identity,
\begin{eqnarray}
\widehat{H}^{R,A}
&\equiv&
\tilde{\varepsilon}^{R,A}(\varepsilon)\whtauz-\widehat\Delta(\hat{\vec{p}}) 
=
C^{R,A}(\vec{p},\varepsilon)\,\widehat{g}_0^{R,A}(\vec{p},\varepsilon)
\,,
\label{eq-H-g_identity}
\\
\mbox{where}\quad C^{R,A}
&\equiv& 
-\frac{1}{\pi}\sqrt{\mid\Delta(\hat{\vec{p}})\mid^2 - (\tilde{\varepsilon}^{R,A}(\varepsilon))^2}
\,.
\label{eq-Denominator}
\end{eqnarray}
The resulting equations for $\delta\widehat{g}^{R,A,a}$ become,
\begin{eqnarray}\label{deltags}
\delta\widehat{g}^{R,A} 
&=&
-\frac{1}{2\pi^2 C^{R,A}}\,\widehat{g}_0^{R,A}\,\left[\widehat{g}_0^{R,A}\,,\,\delta\widehat{\Sigma}^{R,A}\right]
\,,
\\
\delta\widehat{g}^a 
&=& 
\frac{\widehat{g}_0^R}{\pi^2\,C_+^a}
\left[\widehat{g}_0^R-\widehat{g}_0^A\right]
\times
i\vec{v}_{\hat{\vec{p}}}\cdot\vec{\nabla}\phi_0
+
\delta\widehat{\Sigma}^a\,\widehat{g}_0^A-\widehat{g}_0^R\,\delta\widehat{\Sigma}^a
\,,
\label{eq-ga}
\end{eqnarray}
where $C^a_{+}=C^R+C^A$. 

In the IHSM for p-wave pairing we consider pure $s$-wave scattering ($l=0$), in which case the equilibrium T-matrix is diagonal,
\begin{equation}\label{t1t3}
\widehat{T}^{R,A} = T_1^{R,A}\widehat{1} + T_3^{R,A}\whtauz
\,,
\end{equation}
where $T_{1,3}^{R,A}$ are easily obtained in terms of the components of the equilibrium propagator, 
$\widehat{g}_0$, in Eq.~\eqref{eq-H-g_identity} and Eq.~\eqref{t-matrix-eqn} for $\widehat{T}^{R,A}$.
For the chiral A-phase with the orbital order parameter given by $\Delta(\hat{\vec{p}})=\Delta\,\sin\theta_{\hat{\vec{p}}}\,e^{i\phi_{\hat{\vec{p}}}}$, where $\theta_{\hat{\vec{p}}}$ ($\phi_{\hat{\vec{p}}}$) is the polar (azimuthal) angle of $\hat{\vec{p}}$ relative to (in the plane perpendicular to) the chiral axis, $\vec{\ell}$.
The corresponding T-matrix components in the IHSM reduce to
%
%
\begin{eqnarray}\label{t13A}
T_1^{R,A} &=& \frac{\sqrt{\bar\sigma}}{\pi N_f}\,\times
\frac{\sqrt{(1-\bar\sigma)}}
{(1-\bar\sigma)-\bar{\sigma}\,G^{R,A}(\varepsilon)^2}
\,,
\\
T_3^{R,A} &=& \frac{\sqrt{\bar\sigma}}{\pi N_f}\,\times
\frac{\sqrt{\bar\sigma}\,G^{R,A}(\varepsilon)}
{(1-\bar\sigma)-\bar{\sigma}\,G^{R,A}(\varepsilon)^2}
\,,
\end{eqnarray}
where
\begin{equation}\label{GRA}
G^{R,A}(\varepsilon)\equiv 
\left\langle
\frac{\tilde{\varepsilon}^{R,A}(\varepsilon)}
{\sqrt{\Delta^2\sin^2\theta_{\hat{\vec{p}}} - \tilde{\varepsilon}^{R,A}(\varepsilon)^2}}
\right\rangle_{\hat{\vec{p}}}
\,.
\end{equation}
The T-matrix determines the equilibrium sub-gap spectrum shown in Fig.~\ref{figAphaseDos}, and also the vertex correction to the thermal conductivity tensor as shown below. 

Heat current is carried by both quasiparticles and quasiholes, and as a result is proportional to the trace of the Keldysh propagator (c.f. Eq.~\eqref{je}). However, the trace of the local equilibrium retarded and advanced propagators vanish, so that $\Tr{\delta\widehat{g}^{K}}=\Tr{\delta\widehat{g}^a}$.  
Noting also that the driving term, 
\begin{equation}\label{eq-driving_term}
\vec{v}_{\hat{\vec{p}}}\cdot\vec{\nabla}\phi_0 
= 
-\frac{\varepsilon}{2T^2}\mbox{\sf sech}^2\left(\frac{\varepsilon}{2T}\right)\,
\vec{v}_{\hat{\vec{p}}}\cdot\vec{\nabla}T
\,,
\end{equation}
and combining Eqs.~\eqref{eq-ga} and \eqref{eq-driving_term} with Eq.~\eqref{je} we can then express the components of the heat current in terms of the direct (d) and anomalous (vertex correction) terms, $\vec{j}_{\varepsilon}=\vec{j}^d_{\varepsilon}+\vec{j}^a_{\varepsilon}$, with $\vec{j}_{\varepsilon,i}^{d,a} = -K^{d,a}_{ij}\,\nabla_j T$.
The direct term for the heat current reduces to the following contribution to the conductivity tensor,
\begin{eqnarray}\label{k-expression}
K^d_{ij}=\ns
\frac{N_f}{8\pi\,T^2}\ns
\int_{-\infty}^{+\infty}\ns\ns\ns d\varepsilon\,\varepsilon^2
                        \mbox{\sf sech}^2\left(\frac{\varepsilon}{2T}\right)
\int\ns d^2\vec{p}\,
\frac{(\vec{v}_{\hat{\vec{p}}})_i\,(\vec{v}_{\hat{\vec{p}}})_j}
{\pi^2 C_+^a(\hat{\vec{p}},\varepsilon)}
\times
\Tr{\pi^2\widehat{\mathsf{1}} + \widehat{g}_0^R\,\widehat{g}_0^A}
\,,
\end{eqnarray}
which is manifestly diagonal and contributes only to the longitudinal heat current. This term gives
to the result for the thermal conductivity reported in Ref.~\cite{sha03}. 

The vertex correction to the heat current results from the anomalous contribution to the self energy,
\begin{equation}\label{anomalousje}
\vec{j}_{\varepsilon}^a = N_f
\int_{-\infty}^{+\infty}\ns\ns\frac{d\varepsilon}{4\pi i}
\int\ns d^2\vec{p}\,
\frac{\varepsilon\,\vec{v}_{\hat{\vec{p}}}}{\pi^2 C_+^a(\hat{\vec{p}},\varepsilon)}
\Tr{\pi^2\,\delta\widehat{\Sigma}^a + \widehat{g}_0^R\,\delta\widehat{\Sigma}^a\,\widehat{g}_0^A}
\,,
\end{equation}
where the anomalous self-energy is given by,
\begin{equation}\label{sigmaa-tmatrix}
\delta\widehat{\Sigma}^a 
\equiv 
\bar{n}_s\,N_f\,
\langle\widehat{T}^R\,\delta\widehat{g}^a\,\widehat{T}^A\rangle_{\hat{\vec{p}}}
=\bar{n}_s\,N_f
\int\ns d^2\vec{p}\,
\widehat{T}^R\,\delta\widehat{g}^a\,\widehat{T}^A 
\,.
\end{equation}
It is useful to parametrise the equilibrium propagator defined in Eqs.~\eqref{eq-H-g_identity} and \eqref{eq-Denominator} in terms of four Nambu components,
\begin{equation}\label{g0-parametrization}
\widehat{g}_0^{R,A} 
= 
-\pi
\begin{pmatrix}
g^{R,A} & f^{R,A}
\cr
-\underline{f}^{R,A}
&-g^{R,A}
\end{pmatrix}
\,,
\end{equation}
where $\underline{f}^{R,A}=(f^{R,A})^{\star}$. Similarly, we parametrise the anomalous propagator and self energy as
\begin{equation}\label{a-parametrization}
\delta\widehat{g}^{a} 
=
-\pi
\begin{pmatrix}
\delta g		&	\delta f
\cr
-\delta\underline{f}	&	\delta\underline{g}
\end{pmatrix}
\,,\quad
\delta\widehat{\Sigma}^{a}=
\begin{pmatrix}
\delta\Sigma			&	\delta \Delta
\cr
-\delta\underline{\Delta}	&	\delta\underline{\Sigma}
\end{pmatrix}
\,.
\end{equation}
The equations for $\delta\Sigma^a$ in terms of $\delta g^a$ then simplify to
\begin{eqnarray}\label{sa-in-ga-parametrised}
\delta\Sigma 
&=&
-\pi\bar{n}_s N_f\langle\delta{g}\rangle_{\hat{\vec{p}}}\times\,T_+^R\,T_+^A
\\
\delta\underline{\Sigma} 
&=& 
-\pi\bar{n}_s N_f\langle\delta\underline{g}\rangle_{\hat{\vec{p}}}\times\,T_-^R\,T_-^A
\\
\delta\Delta 
&=&
-\pi\bar{n}_s N_f\langle\delta{f}\rangle_{\hat{\vec{p}}}\times\,T_+^R\,T_-^A
\label{eq-anomalous-Delta}
\\
\delta\underline{\Delta} 
&=& 
-\pi\bar{n}_s N_f\langle\delta\underline{f}\rangle_{\hat{\vec{p}}}\times\,T_-^R\,T_+^A
\,.
\label{eq-anomalous-Deltabar}
\end{eqnarray}
where $T^{R,A}_{\pm}=T^{R,A}_1 \pm T^{R,A}_3$. Then with the short-hand notation for $\mathscr{C}^a=-\nicefrac{1}{C_+^a}$ we can express the components of the anamolous propagator as
\begin{equation}\label{gamatrixeqn}
\begin{pmatrix}
\delta g
\cr
\delta\underline{g}
\cr
\delta f
\cr
\delta\underline{f}
\end{pmatrix}
=
\mathcal{L}^a\,\times
\left\{
\begin{pmatrix} 1\cr 1 \cr 0 \cr 0 \end{pmatrix} \cdot (i\vec{v}_{\hat{\vec{p}}}\cdot\vec{\nabla}\phi_0)
+
\begin{pmatrix}
\delta\Sigma \cr \delta\underline{\Sigma} \cr \delta\Delta \cr \delta\underline{\Delta} \end{pmatrix}
\right\}
\,,
\end{equation}
where 
\begin{eqnarray}\label{La}
\mathcal{L}^a 
&=& 
\mathscr{C}^a
\begingroup
\renewcommand*{\arraystretch}{1.25}
\begin{pmatrix}
1 + g^R g^A 	
\quad
& -f^R f^A	
\quad
& -g^R \underline{f}^A 
\quad
& -f^R g^A
\cr
-f^R f^A	
& 1 + g^R g^A 	
\quad
& \underline{f}^R g^A 
\quad
& g^R f^A
\cr
g^R f^A 
\quad
& -f^R g^A 
\quad
& 1 - g^R g^A 	
\quad
& -f^R\,f^A
\cr
\underline{f}^R g^A 
\quad
& -g^R \underline{f}^A 
\quad
& -\underline{f}^R \underline{f}^A
\quad
& 1 - g^R g^A 	
\end{pmatrix}
\endgroup
\,,
\end{eqnarray}
For the chiral ESP state considered here we can simplify the matrix elements using the identities, 
$g^R = (g^A)^{\star}$ and $f^R=(f^A)^{\star}$ in which case 
$g^R g^A = \mid g^R\mid^2 = \mid g^A \mid^2 \equiv \mid g \mid^2$,
and similarly, 
$f^R f^A = \mid f^R\mid^2 = \mid f^A \mid^2 \equiv \mid f \mid^2$.
We also obtain
\begin{equation}
\mathscr{C}^a = \frac{\pi}{2}
\frac{1}{\mbox{\sf Re}\sqrt{\mid\Delta(\hat{\vec{p}})\mid^2-\tilde{\varepsilon}^R(\varepsilon)^2}}
\,.
\end{equation}
The terms from the first column vector in Eq.~\eqref{gamatrixeqn} generates the direct contribution to the thermal conductivity, i.e. the result in Eq.~\eqref{k-expression}. 

In order to calculate the vertex correction to the heat current we need to evaluate Fermi surface averages of the anomalous pair propagators, $\delta f$ and $\delta \underline{f}$. The driving term in Eq.~\eqref{gamatrixeqn}, and the matrix $\mathcal{L}^a$, dictates their momentum dependences: 
$\delta f\propto\Delta(\hat{\vec{p}})\,\vec{v}_{\hat{\vec{p}}}\cdot\vec{\nabla}T$ and 
$\delta \underline{f}\propto\underline{\Delta}(\hat{\vec{p}})\,\vec{v}_{\hat{\vec{p}}}\cdot\vec{\nabla}T$. 
We write $\delta f = \delta f_j\,\nabla_j T$ and similarly, $\delta \underline{f} = \delta \underline{f}_j\,\nabla_j T$, and since the direction $\vec{\nabla}T$ is fixed, but arbitrary, we can express the angular averages of the Cartesian components of the anomalous pair propagators as
\begin{eqnarray}
\langle\delta{f}_j\rangle_{\hat{\vec{p}}} 
&=& 
-i\frac{\varepsilon}{2T^2}\,\mbox{\sf sech}^2\left(\frac{\varepsilon}{2T}\right)\,
\langle (\vec{v}_{\hat{\vec{p}}})_j\,\mathscr{C}^a(g^R f^A - f^R g^A)\rangle_{\hat{\vec{p}}} 
\nonumber\\
&+&
\langle \mathscr{C}^a\,(1-\ns\mid g\mid^2)\rangle_{\hat{\vec{p}}}\, 
\delta\Delta_j 
\,,
\label{eq-anomalous-f_average1}
\\
\langle\delta\underline{f}_j\rangle_{\hat{\vec{p}}} 
&=& 
-i\frac{\varepsilon}{2T^2}\,\mbox{\sf sech}^2\left(\frac{\varepsilon}{2T}\right)\,
\langle (\vec{v}_{\hat{\vec{p}}})_j\,\mathscr{C}^a(\underline{f}^R g^A - g^R \underline{f}^A)\rangle_{\hat{\vec{p}}} 
\nonumber\\
&+&
\langle \mathscr{C}^a\,(1-\ns\mid g\mid^2)\rangle_{\hat{\vec{p}}}\, 
\delta\underline{\Delta}_j
\,. 
\label{eq-anomalous-f_average2}
\end{eqnarray}
The two angular averages involving the Fermi velocity reduce to
\begin{eqnarray}
\langle (\vec{v}_{\hat{\vec{p}}})_j\,\mathscr{C}^a(g^R f^A - f^R g^A)\rangle_{\hat{\vec{p}}} 
&=&
-(\tilde{\varepsilon}^R(\varepsilon) - \tilde{\varepsilon}^A(\varepsilon))
\label{eq-vertex_average1}
\\
&\times&
\left\langle
\frac{\mathscr{C}^a\,(\vec{v}_{\hat{\vec{p}}})_j\,\Delta(\hat{\vec{p}})}
     {\sqrt{\mid\Delta(\hat{\vec{p}})\mid^2-\tilde{\varepsilon}^R(\varepsilon)^2}
      \sqrt{\mid\Delta(\hat{\vec{p}})\mid^2-\tilde{\varepsilon}^A(\varepsilon)^2}}
\right\rangle_{\hat{\vec{p}}} 
\,,
\nonumber
\\
\langle (\vec{v}_{\hat{\vec{p}}})_j\,\mathscr{C}^a(\underline{f}^R g^A - g^R \underline{f}^A)\rangle_{\hat{\vec{p}}} 
&=&
+(\tilde{\varepsilon}^R(\varepsilon) - \tilde{\varepsilon}^A(\varepsilon))
\label{eq-vertex_average2}
\\
&\times&
\left\langle
\frac{\mathscr{C}^a\,(\vec{v}_{\hat{\vec{p}}})_j\,\Delta^*(\hat{\vec{p}})}
     {\sqrt{\mid\Delta(\hat{\vec{p}})\mid^2-\tilde{\varepsilon}^R(\varepsilon)^2}
      \sqrt{\mid\Delta(\hat{\vec{p}})\mid^2-\tilde{\varepsilon}^A(\varepsilon)^2}}
\right\rangle_{\hat{\vec{p}}} 
\nonumber
\,.
\end{eqnarray}
Using Eqs.~\eqref{eq-vertex_average1} and \eqref{eq-vertex_average2} with Eqs.~\eqref{eq-anomalous-f_average1} and \eqref{eq-anomalous-f_average2} to evaluate the anomalous self energies, Eq.~\eqref{eq-anomalous-Delta} and ~\eqref{eq-anomalous-Deltabar} we obtain the solutions,
\begin{eqnarray}\label{dDs}
\hspace*{-9mm}
\delta\Delta_j 
\ns&=&\ns
-\frac{\pi\bar{n}_s N_f T_+^R\,T_-^A}{1 - \mathcal{A}}
\left(\frac{i\varepsilon}{2T^2}\right)
\mbox{\sf sech}^2\left(\frac{\varepsilon}{2T}\right)
\langle\langle\,(\vec{v}_{\hat{\vec{p}}})_j\,\Delta(\hat{\vec{p}})\rangle\rangle\,
(\tilde{\varepsilon}^R - \tilde{\varepsilon}^A)
\,,
\\
\hspace*{-9mm}
\delta\underline\Delta_j 
\ns&=&\ns
+\frac{\pi\bar{n}_s N_f T_-^R\,T_+^A}{1 - \mathcal{A}^*}
\left(\frac{i\varepsilon}{2T^2}\right)
\mbox{\sf sech}^2\left(\frac{\varepsilon}{2T}\right)
\langle\langle\,(\vec{v}_{\hat{\vec{p}}})_j\,\Delta^*(\hat{\vec{p}})\rangle\rangle\,
(\tilde{\varepsilon}^R - \tilde{\varepsilon}^A)
\,.
\end{eqnarray}
\vspace*{-8mm}
\begin{eqnarray}
\mbox{where}\qquad
\mathcal{A}
&\equiv& 
-\pi\,\bar{n}_s\,N_f\,T_+^R\,T_-^A\,\left\langle\mathscr{C}^a(1 - \vert g\vert^2)\right\rangle_{\hat{\vec{p}}} 
\,,
\label{A-defn}
\\
\mbox{and}\quad
\langle\langle \ldots \rangle\rangle 
&\equiv& 
\left\langle
\frac{\mathscr{C}^a(\hat{\vec{p}},\varepsilon)\,(...)}
     {\sqrt{\vert\Delta(\hat{\vec{p}})\vert^2 - \tilde{\varepsilon}^R(\varepsilon)^2}
      \sqrt{\vert\Delta(\hat{\vec{p}})\vert^2 - \tilde{\varepsilon}^A(\varepsilon)^2}}
\right\rangle_{\hat{\vec{p}}} 
\,.
\label{AngAvg-defns}
\end{eqnarray}
These nonequilibrium pair potentials (vertex corrections) define the anomalous contribution to the conductivity tensor,
\begin{eqnarray}\label{k-A}
\hspace*{-5mm}
K_{ij}^a
\ns=\ns
-N_f\ns\int\ns\frac{d\varepsilon}{4\pi i}\varepsilon
\left\langle
\mathscr{C}^a(\vec{v}_{\hat{\vec{p}}})_i
\ns
\left[
(\underline{f}^R g^A\ns-\ns g^R \underline{f}^A)\delta\Delta_j
\ns+\ns
(g^R f^A\ns-\ns f^R g^A)\delta\underline{\Delta}_j
\right]
\right\rangle_{\hat{\vec{p}}}
\ns.
\end{eqnarray}
A key observation is the following: 
\begin{eqnarray}\label{sin2theta}
\langle\langle\vec{v}_{\hat{\vec{p}}}^x\,\Delta(\hat{\vec{p}})\rangle\rangle 
&=& 
\langle\langle\vec{v}_{\hat{\vec{p}}}^x\,\Delta^{\star}(\hat{\vec{p}})\rangle\rangle 
= 
\frac{v_f\,\Delta}{2}\langle\langle\sin^2\theta_{\hat{\vec{p}}}\rangle\rangle 
\,,
\\
\langle\langle\vec{v}_{\hat{\vec{p}}}^y\,\Delta(\hat{\vec{p}})\rangle\rangle 
&=& 
-\langle\langle\vec{v}_{\hat{\vec{p}}}^y\,\Delta^{\star}(\hat{\vec{p}})\rangle\rangle 
= 
i\frac{v_f\,\Delta}{2}\langle\langle\sin^2\theta_{\hat{\vec{p}}}\rangle\rangle 
\,.
\end{eqnarray} 
These identities, combined with Eqs.~\eqref{k-A},~\eqref{A-defn2} and~\eqref{B-defn},
and a simple rescaling of the T-matrix components,
\begin{equation}
\bar{T}_{\pm}^{R,A} = \frac{\pi N_f}{\sqrt{\bar\sigma}}\,T_{\pm}^{R,A}
\,,
\end{equation}
reduce to the result for the anoamolous Hall conductivity, $K_{xy}^a$ in Eq.~\eqref{Kaij} with
\begin{equation}\label{A-defn2}
\mathcal{A} \equiv -\Gamma_N\,\bar{T}_+^R \bar{T}_-^A
\left\langle
\mathscr{C}^a\,(1 - \vert g \vert^2)
\right\rangle_{\hat{\vec{p}}}
\,,
\end{equation}
\begin{equation}\label{B-defn}
\mbox{and}\quad
\mathcal{B}\equiv\langle\langle \sin^2\theta_{\hat{\vec{p}}}\rangle\rangle
=
\left\langle
\frac{\mathscr{C}^a(\hat{\vec{p}},\varepsilon)\sin^2\theta_{\hat{\vec{p}}}}
     {\sqrt{\vert\Delta(\hat{\vec{p}})\vert^2 - \tilde{\varepsilon}^R(\varepsilon)^2}
      \sqrt{\vert\Delta(\hat{\vec{p}})\vert^2 - \tilde{\varepsilon}^A(\varepsilon)^2}}
\right\rangle_{\hat{\vec{p}}} 
\,.
\end{equation}
The vertex corrections to longitudinal components of the thermal conductivity follow similarly to yield Eq.~\eqref{Kaij} 
for $K^a_{xx}$. 

\end{appendices}


\end{document}